\documentstyle[prd,eqsecnum,aps]{revtex}

\newcommand{\beq}{\begin{equation}} \newcommand{\eeq}{\end{equation}}
\def\bea{\begin{eqnarray}} \def\eea{\end{eqnarray}}
\newcommand{\ba}{\begin{array}} \newcommand{\ea}{\end{array}}
\def\a{\alpha}   \def\G{\Gamma}
\def\d{\delta}  \def\ee{\varepsilon} 
 \def\th{\theta}  
    
  \def\p{\pi}  
  \def\t{\tau} 
    
\def\Ps{\Psi} \def\o{\omega} \def\O{\Omega} \def\vf{\varphi} \def\vr{\varrho}
\def\pa{\partial}  
\def\sqr#1#2{{\vcenter{\hrule height.#2pt \hbox{\vrule width.#2pt
height#1pt \kern#1pt \vrule width.#2pt} \hrule height.#2pt}}}

\def\IR{{I\kern-0.25em R}}

\begin{document}

\draft

\title{ Spatial Geometry of the Electric Field Representation of
\\ Non-Abelian Gauge Theories\footnote{Research supported in part by
NSF Grant PHY/92-60867, Ministerio de Educaci\'on y Ciencia, Spain,
and by the U.S. Department of Energy (D.O.E.)
under contract \#DE-FC02-94ER40818.
\\ E-mail: {\tt mbauer@surya11.cern.ch, freedman@surya11.cern.ch,
hagensen@ebubecm1}.}} \author{Michel Bauer\footnote{Permanent address:
S.Ph.T.~-- Saclay, 91191 Gif-sur-Yvette CEDEX, France.}, Daniel Z.
Freedman\footnote{Permanent address: Department of Mathematics and
Center for Theoretical Physics, Massachusetts Institute of Technology,
Cambridge, MA, 02139, USA.}} \address{CERN -- TH-Division, CH-1211
Gen\`eve 23, SWITZERLAND}

\author{and}

\author{Peter E. Haagensen}
\address{Departament d'Estructura i Constituents de la Mat\`eria,
\\Facultat de F\'\i sica, Universitat de Barcelona\\ Diagonal, 647\ \
08028 Barcelona, SPAIN}\bigskip

\maketitle

\bigskip
\begin{abstract}

A unitary transformation $\Ps [E]=\exp (i\O [E]/g) F[E]$ is used to
simplify the Gauss law constraint of non-abelian gauge theories in the
electric field representation. This leads to an unexpected
geometrization because $\o^a_i\equiv -\d\O [E]/\d E^{ai}$ transforms as
a (composite) connection. The geometric information in $\o^a_i$ is
transferred to a gauge invariant spatial connection $\G^i_{jk}$ and
torsion by a suitable choice of basis vectors for the adjoint
representation which are constructed from the electric field $E^{ai}$.
A metric is also constructed from $E^{ai}$. For gauge group $SU(2)$, the
spatial geometry is the standard Riemannian geometry of a 3-manifold,
and for $SU(3)$ it is a metric preserving geometry with both
conventional and unconventional torsion. The transformed Hamiltonian is
local. For a broad class of physical states, it can be expressed
entirely in terms of spatial geometric, gauge invariant variables.

\end{abstract}
\vspace{1cm}
\pacs{CERN--TH. 7238/94\ \ \ April 1994}

\section{introduction}\bigskip

The canonical commutation relations and Gauss law constraint of
Hamiltonian gauge theories in temporal gauge are invariant under spatial
diffeomorphisms of the canonical variables $A^a_i(x)$ and $E^{ai}(x)$.
This local $GL(3)$ symmetry is broken in the Hamiltonian in a simple way
because of the appearance of the Cartesian metric $\d_{ij}$ of flat
space, and the energy density transforms as a $GL(3)$ tensor density.
In this paper we discuss a formulation of non-abelian gauge theories in
which the Gauss law constraint is easily implemented and the Hamiltonian
is expressed in terms of variables which are gauge invariant or
covariant and also geometric, i.e. they are $GL(3)$ tensors, connections
or curvatures.  The resulting theory has an elegant mathematical
structure but it is far from clear that the spatial geometry will be
helpful for dynamical calculations or offer any advantages over such
well-developed approaches as lattice gauge theories.

We choose to work in the electric representation of gauge theories in
which states $\Psi[E^{ai}]$ are functionals of the electric field. In
common with an earlier non-geometric approach to the $SU(2)$ theory
\cite{gj} the key element of our work is a unitary transformation
$\Psi[E]=\exp(i\O [E]/g) F[E]$ of the theory which simplifies the form
of the Gauss law constraint. The phase $\O [E]$ is a local $GL(3)$
invariant functional of the electric field, whose variation under
infinitesimal gauge transformations is $\d\O [E] =\int d^3\!x\ \
\th^a\pa_iE^{ai}$. These gauge and $GL(3)$ properties of $\O [E]$ imply
that the quantity $\o^a_i[E]=-\d\O [E]/\d E^{ai}$ transforms as a Lie
algebra valued connection on the initial value surface $\IR^3$. Thus a
composite gauge connection $\o^a_i[E]$ appears and plays a central role
in our formulation although the fundamental variable $E^{ai}$ transforms
homogeneously under gauge transformations. The Hamiltonian is local
but, as in earlier work \cite{gj,iksf} it involves functional
derivatives $\d/\d E^{ai}$ up to fourth order.

For gauge group $SU(2)$, $\o^a_i[E]$ is simply the standard Riemannian
spin connection on a three-manifold with frame 1-form $e^a_i(x)$ related
to the electric field by $E^{ai}=\ee^{abc} \ee^{ijk} e^b_je^c_k/2$. One
can argue that under fairly general assumptions one can restrict to wave
functionals $F[G_{ij},\vr ]$ where $G_{ij}=e^a_i e^a_j$ is a composite
metric and $\vr=\det E^{ai}$. Such states satisfy the Gauss law
constraint, and the Hamiltonian acting on them can be rewritten in terms
of the Christoffel connection $\G^k_{ij}$ and curvature $R^k_{\;\; \ell
ij}$.  Thus a Riemannian spatial geometry underlies $SU(2)$ gauge
theory.

It is actually known \cite{hns,nm} from work on Ashtekhar
variables in gravity that the spin connection on a 3-manifold is the
variational derivative of the local functional $\O [E]$.  It is not lost
upon us that the Ashtekhar approach makes gravity look a lot like gauge
theory, while our approach makes gauge theory look a lot like gravity.

One could view the structure described above as the accidental
consequence of the fact that the gauge group $SU(2)$ co\"{\i}ncides with
the tangent space group of a three-manifold.  However we are able to
give a formula for the phase $\O [E]$ for a general gauge group $G$.
The formula is not entirely explicit because it involves the inverse of
a matrix of dimension $3\dim G \times 3\dim G$ which is a quadratic
function of $E^{ai}$.  But it is explicit enough to see that the general
structure of the theory is similar to $SU(2)$, but that the associated
spatial geometry, which we outline for $SU(3)$, is more complicated.  It
can be described as a metric-preserving geometry with an unconventional
torsion.

One may also study the spatial geometry of a magnetic formulation of
gauge theory.  Indeed we drew our inspiration from a recent study
\cite{fhjl} of the $SU(2)$ theory in which a curious Einstein space
geometry with torsion appeared.  The geometry is correct, but the
application made to Hamiltonian dynamics in \cite{fhjl} failed because
of the Wu-Yang ambiguity \cite{wy} which is generically continuous in
three spatial dimensions \cite{fk}.  A new magnetic formulation
\cite{hj} avoids the problem and leads to a Hamiltonian which is second
order in functional derivatives $\d/\d G_{ij}$ with respect to a
composite metric variable, but is non-local.

We also wish to cite recent papers involving a geometrical approach to
gauge theories in the Lagrangian formalism by Lunev \cite{l} and others
\cite{h,ns} in which a spatial metric has appeared in studies of gauge
theories.  Finally, there are recent extensive studies of Hamiltonian
dynamics for gauge theory in light-cone gauge \cite{pw}.\bigskip

\section{the unitary transformation and its consequences}

The canonical variables of a non-abelian gauge theory are the vector
potential $A^a_i(x)$ and electric field $E^{ai}(x)$ which satisfy the
commutation relations
\beq \label{eq commut} \left[A^a_i(x),E^{bj}(x')\right]=i\d^{ab} \d^j_i
\d^{(3)}(x-x'). \eeq
In temporal gauge, $A^a_0(x)=0$, the generator of spatial gauge
transformations with parameter $\th^a(x)$ is
\beq\label{eq gau1} \begin{array}{l} {\cal G}[\th ]=\int d^3\!x\
\th^a(x)\ {\cal G}^a(x) \\*[8pt] \displaystyle{ {\cal G}^a(x)={1\over
g}\ D_iE^{ai}(x)=\frac{1}{g}\left( \pa_iE^{ai}(x)+g\,f^{abc}A^b_i(x)
E^{ci}(x)\right),} \end{array} \eeq
and Eq.\ (\ref{eq commut}) implies the quantum gauge transformation
rules
\beq\label{eq gau} \begin{array}{l} \d A^a_i(x)=-i \left[{\cal G}[\th
],A^a_i(x) \right]= \displaystyle{ \frac{1}{g} \left(\pa_i \th^a(x)
+g\,f^{abc}A^b_i(x)\th^c(x)\right)} \\*[8pt] \d E^{ai}(x)=-i\left[{\cal
G}[\th ],E^{ai}(x)\right]=f^{abc} E^{bi}(x)\th^c(x). \end{array}\eeq
Using the magnetic field
\beq\label{eq B} B^{ai}(x)=\ee^{ijk} \left(\pa_jA^a_k(x) +\frac{1}{2}\
gf^{abc} A^b_j(x)A^c_k(x) \right) \eeq
which transforms homogeneously, i.e. as the electric field in
Eq.~(\ref{eq gau}), the Hamiltonian can be written as
\beq H=\frac{1}{2}\int d^3\!x\ \d_{ij}\Bigl(
E^{ai}(x)E^{aj}(x)+B^{ai}(x)B^{aj}(x)\Bigr) \ . \eeq

We now observe that Eqs.~(\ref{eq commut}-\ref{eq B}) are covariant
under coordinate transformations $x^i \to y^{\a}$ on the domain $\IR^3$
provided that

\begin{enumerate}
\item $A^a_i(x)$ transforms as a covariant vector
\beq A'^a_{\a}(y)=\frac{\pa x^i}{\pa y^{\a}} A^a_i(x) \eeq
which is implied by the 1-form interpretation ${\bf A}^a=A^a_i dx^i$ of
the vector potential and
\item $E^{ai}(x)$ transforms as a contravariant vector density
\beq E'^{a\a}(y)=\left|\frac{\pa x}{\pa y}\right| \frac{\pa y^{\a}}{\pa
x^i} E^{ai}(x)\eeq
which is consistent with its implementation as a functional derivative
$E^{ai}(x)=-i \delta/\delta A^a_i(x)$ in the familiar magnetic
representation of (\ref{eq commut}).
\end{enumerate}
Note that the gauge parameters $\th^a(x)$ transform as $GL(3)$ scalars
and that ${\cal G}^a(x)$ is a scalar density. No connection $\G^i_{jk}$
is required in Eq.~(\ref{eq gau1}) because $E^{ai}$ is a density of
weight one. The magnetic field is also a contravariant vector density
of weight one.

The Hamiltonian fails to be $GL(3)$ invariant because the fixed
cartesian metric appears, but one sees that the energy density
transforms as the $\d_{ij}$ trace of a contravariant symmetric tensor
density of weight two. The Hamiltonian is gauge invariant, viz.,
$\left[{\cal G}[\th ],H\right]=0$, and the dynamical problem of gauge
theories can be formally stated as the problem of diagonalizing $H$ on
the physical subspace of gauge invariant states $\Psi$ which satisfy the
Gauss law constraint
\beq {\cal G}^a(x) \Psi =\frac{1}{g}\, ( \pa_iE^{ai}(x) +g\,f^{abc}
A^b_i(x)E^{ci}(x)) \Psi=0\ . \eeq
Our goal here is to formulate this dynamical problem in a way which
maintains the $GL(3)$ properties of the theory.

We work in electric field representation with state functionals
$\Psi[E]$. Then $E^{ai}(x)$ is realized by simple multiplication and
$A^a_i(x)=i\delta / \delta E^{ai}(x)$ by functional differentiation.
It would be easy to implement the Gauss law constraint if the gauge
generator contained only the second term
\beq\bar{{\cal G}}^a(x)=-if^{abc}E^{bi}(x)\frac{\d }{\d
E^{ci}(x)} \eeq
because this operator simply generates local group rotations without
spatial transport. Note that both ${\cal G}^a(x)$ and $\bar{{\cal
G}}^a(x)$ satisfy the group algebra in the local form
\beq \left[ {\cal G}^a(x),{\cal G}^b(x')\right]=i f^{abc}\d^{(3)}
(x-x')\, {\cal G}^c(x).\eeq

In the spirit of \cite{gj}, we shall make a unitary transformation on
the states and operators of the theory in order to simplify the gauge
generators. We write
\beq \begin{array}{l} \label{eq phas} \Psi[E]=\exp(i\O [E]/g)\ F[E]
\\*[6pt] {\cal O}(x)=\exp(i\O [E]/g)\ \bar{\cal O}(x)\ \exp(-i\O [E]/g)
\end{array}\eeq
and require that
\beq\label{eq gphas} {\cal G}^a(x)\ \exp(i\O [E]/g)\ F[E]= \exp(i\O
[E]/g)\ \bar{{\cal G}}^a(x)\ F[E]. \eeq
The phase $\O [E]$ thus satisfies
\beq\begin{array}{lcl} \exp(-i\O [E]/g)\ {\cal G}^a(x)\ \exp(i\O [E]/g)
& = & \displaystyle{ {\cal G}^a(x)+{i\over g}\left[{\cal G}^a(x),\O
[E]\ \right]} \\*[8pt] & = & \bar{{\cal G}}^a(x) \end{array}\eeq
This is equivalent to the requirement that the gauge variation of
$\O [E]$ be
\beq \label{eq varo} \delta \Omega[E] =-i\ \left[\, {\cal G}[\th ]\,
,\Omega[E]\, \right] =\ \int d^3\!x\ \th^a(x)\pa_iE^{ai}(x)\ .\eeq
We also require that the phase $\Omega[E]$ be $GL(3)$ invariant, so that
the unitary transformation preserves the behavior of the theory under
spatial diffeomorphisms. Note that for an abelian gauge group $U(1)$
any $\O [E]$ is gauge invariant, so that we cannot satisfy Eq.~(\ref{eq
varo}). Thus our treatment must be restricted to non-abelian groups. A
unitary transformation of similar structure appears in a recent study of
a $1+1$ dimensional gravity theory \cite{cj}.

We will now show that the form of the resulting theory is essentially
determined by these two requirements on $\O [E]$. In subsequent
sections we will give local formulae for the phase, i.e. of the form $\O
[E]=\int d^3\!x\ f(E(x),\pa E(x))$ first for gauge group $SU(2)$ and
then for general $G$.

So we now assume the existence of a $GL(3)$ invariant phase whose gauge
variation is given by Eq.~(\ref{eq varo}), and work out the structure of
the unitary transformed theory. The transformed canonical variables are
\beq \begin{array}{lcl} \bar{E}^{ai}(x) & = & \exp(-i\O [E]/g)\
E^{ai}(x)\ \exp(i\O [E]/g)=E^{ai}(x) \\*[8pt] \bar{A}^a_i(x) & = &
\exp(-i\O [E]/g)\ A^a_i(x)\ \exp(i\O [E]/g) \\*[8pt] & = &
\displaystyle{ A^a_i(x) +{i\over g}\left[A^a_i(x),\O [E]\
\right]}\\*[8pt] & = & \displaystyle{ A^a_i(x)-\frac{1}{g} \frac{\d}{\d
E^{ai}(x)} \O [E]} \\*[8pt] & \equiv & \displaystyle{ i\frac{\d}{\d
E^{ai}(x)} +\frac{1}{g}\ \o^a_i(x)}\end{array} \label{eq transfo}\eeq
The quantity $\o^a_i(x)$ is the variational derivative of a $GL(3)$
invariant functional with respect to a vector density so $\o^a_i(x)$ is
a covariant vector under spatial diffeomorphisms. Its gauge variation
is

\beq\begin{array}{lcl} \d\o^a_i(x)/g & = & i \left[{\cal G}[\th ],
\frac{\d\O [E]}{\d E^{ai}(x)} \right]/g \\*[6pt] & = & \left[{\cal
G}[\th ],\left[\O [E], A^a_i(x)\right]\ \right] \\*[6pt] & = & \left[\O
[E],\left[{\cal G}[\th ], A^a_i(x)\right]\ \right] - \left[\ \left[{\cal
G}[\th ],\O [E]\ \right], A^a_i(x) \right] \\*[6pt] & = & -i\left[ \O
[E], D_i\th^a(x) \right]/g-i\left[ \int d^3\!y\ \th^b(y) \pa_j
E^{bj}(y), A^a_i(x) \right] \\*[6pt] & = & \displaystyle{ -f^{abc}
\frac{\d}{\d E^{bi}(x)} \O [E]\ \th^c(x)+\frac{1}{g}\ \pa_i \th^a(x)}
\\*[8pt] & = & \displaystyle{ \frac{1}{g}\ \left(\pa_i \th^a(x)+f^{abc}
\o^b_i(x) \th^c(x) \right) \equiv \frac{1}{g}\ \hat{D}_i \th^a
(x)}\end{array}\label{eq delom}\eeq

Thus $\o^a_i[E]/g$ is a local composite function of $E^{ai}$ which
transforms as a gauge potential. One could almost derive this result by
inspection of Eq.~(\ref{eq transfo}), since the gauge variation of
$\bar{A}^a_i$ is
\beq \d \bar{A}^a_i=-i \left[\bar{{\cal G}}[\th],\bar{A}^a_i\right]=
\frac{1} {g} \left(\pa_i \th^a +g\,f^{abc} \bar{A}^b_i \th^c \right).
\eeq
Since $i\d/\d E$ transforms homogeneously, the second term in
Eq.~(\ref{eq transfo}), $\o/g$, must transform as a potential. However
the longer derivation in Eq.~(\ref{eq delom}) has the virtue of
emphasizing that if the gauge variation of any functional $\O[E]$
is given by Eq.~(\ref{eq varo}) then $\d\O/\d E^{ai}$ transforms as a
gauge connection.

In the unitary transformed theory, $\hat{D}_i$ will denote a gauge
covariant derivative formed with the composite connection $\o^a_i$. The
magnetic field formed using Eq.~(\ref{eq B}) with $A$ replaced by $\o/g$
and removing a factor $1/g$ will be denoted by $\hat{B}^{ai}=\ee^{ijk}
\left(\pa_j \o^a_k +\frac{1}{2} f^{abc} \o^b_j \o^c_k \right)$.
It also follows from the trivial relation
\beq \label{triv} \d \O [E] = \int d^3\!x\ \frac{\d \O [E]}{\d E^{ai}}
\d E^{ai} \eeq
and use of Eq.~(\ref{eq varo}) with $\d E^{ai} = f^{abc} E^{bi} \th ^c $
that a ``Bianchi identity'' holds in the form
\beq \label{Bian} \hat{D}_i E^{ai} =0. \eeq

The transformed Hamiltonian is
\beq \label{eq Hbar}\bar{H}=\frac{1}{2}\int d^3\!x\
\d_{ij}(\bar{E}^{ai}\bar{E}^{aj}+ \bar{B}^{ai}\bar{B}^{aj} ).\eeq
The electric term is quite simple. One can define the gauge invariant
symmetric tensor variable
\beq \label{eq phi} \vf^{ij}=E^{ai}E^{aj}\eeq
and express the electric energy density as the multiplication operator
$\frac{1}{2}\d_{ij}\vf^{ij}$.

The magnetic field $\bar{B}^{ai}(x)$ applied to a state $F[E]$ is
\beq\begin{array}{rl} \bar{B}^{ai}F[E] = & \ee^{ijk}
\left(\pa_j\bar{A}^a_k+\frac{1}{2}g \, f^{abc} \bar{A}^b_j \bar{A}^c_k
\right)F[E] \\*[8pt] = & \displaystyle{\left[ \frac{1}{g}\ \hat{B}^{ai}
\right. +i\ \ee^{ijk}\hat{D}_j\frac{\d}{\d E^{ak}} -\frac{g}{2}\
\ee^{ijk}f^{abc} \frac{\d}{\d E^{bj}}\frac{\d}{\d E^{ck}} }\\*[8pt]
&\displaystyle{ \left. \phantom{xxxxxxx}+ i\, g\, \ee^{ijk}f^{abc}
\frac{\d\o^c_k} {\d E^{bj}} \right]F[E] }\end{array} \label{eq Bbar}\eeq

The beginning of a geometric structure is evident in the first two
terms, namely the composite magnetic field and the $\o$-covariant
derivative of $\d F/\d E$. The third term contains the second
functional derivative $\d ^2/\d E \d E$ which is characteristic of the
electric representation of non-abelian theories \cite{gj,iksf}. The
Hamiltonian therefore contains terms up to fourth order in $\d /\d E$.
The fourth term in Eq.~(\ref{eq Bbar}) comes from the operator
reordering $\left[\d / \d E^{bj}(x), \o^c_k(x) \right]$ which was
necessary to obtain the the $\hat{D}_j$ covariant derivative. As will
be seen explicitly for the $SU(2)$ case, this ordering term involves the
singular objects $\pa \d (0)$ and $\d (0)$ and is one troublesome
feature of a nonlinear theory with functional derivatives. Similar
terms also were present in \cite{gj}. Our derivation of the Hamiltonian
has been rather formal and requires regularization. We shall argue in
the appendix that this particular ordering term vanishes if covariant
point splitting regularization is used, but one must study the
additional ordering terms in the magnetic energy density which is
quadratic in $\bar{B}$.

We will discuss the Hamiltonian further in later sections, after we
elucidate its spatial geometric structure. We close this section with a
remark concerning the uniqueness of $GL(3)$-invariant functionals ,which
satisfy Eq.~(\ref{eq varo}). One must not expect a unique solution for
a given gauge group , but the difference $\O '[E] -\O [E]$ between any
two functionals which satisfy the requirements must be both gauge and
$GL(3)$-invariant. For example, one could have
\beq \label{eq diff} \O '[E] - \O [E] \propto \int d^3 x
(\det \vf ^{ij})^{1/4}. \eeq\bigskip

\section{the $SU(2)$ theory}\bigskip

In this section we study the $SU(2)$ gauge theory in more detail. We
first give explicit formulae for the phase $\O [E]$ and composite gauge
connection $\o^a_i$ and then develop the associated spatial geometry
which turns out to be the standard Riemannian geometry of a 3-manifold.

The simplest phase candidate one can write using the electric field
$E^{ai}$ and its matrix inverse $E^a_i$, i.e. $E^{ai}E^b_i=\d^{ab}$,
turns out to be successful. This is
\beq\label{omegasu2} \O [E]\ =\ {1\over2}\int d^3\!x\ \ee^{abc}\
E^{ai}(x)E^{bj}(x)\pa_iE^c_j(x)\ . \eeq
It is $GL(3)$ invariant because the integrand has density weight $+1$
and terms arising from the $\pa_i$ derivative of the co\"ordinate change
of $E^c_j$, which is a covariant vector density, cancel. Although we
need only the infinitesimal gauge variation to confirm Eq.\ (\ref{eq
varo}), it is no more difficult to study the finite gauge transformation
$E^{ai}\to T^{ab}E^{bi}$ where $T^{ab}$ is an $SO(3)$ matrix. We have
\bea\label{gtrsf} \O [TE]\ &=&\ {1\over 2}\int d^3\!x\ \ee^{abc}\
T^{a\bar{a}}T^{b\bar{b}} \left\{ T^{c\bar{c}}E^{\bar{a}i} E^{\bar{b}
j}\pa_iE^{\bar{c}}_j+ \pa_iT^{c\bar{b}}E^{\bar{a}i}\right\}\cr &=&\ \O
[E]-{1\over2}\int d^3\!x\ \ee^{abc}\ (T^{-1}\pa_iT)^{bc}E^{ai}\ .\eea
Group invariance of the structure constants was used to obtain the first
term, and the invariant 1-forms $T^{-1}\pa T$ appear in the second term,
whose infinitesimal limit is Eq.\ (\ref{eq varo}).

We already know that $\o^a_i=-\d \O/\d E^{ai}$ is an $SO(3)$ gauge
connection, so it should not be a great surprise that it turns out to be
a familiar object. We define a new variable $e^a_i$ by
\beq\label{littlee} E^{ai}\ =\ {1\over2}\ee^{ijk}\ee^{abc}e^b_je^c_k\ ,
\eeq
so that $e^a_i$ has dimension $+1$, and is a gauge covariant, $GL(3)$
vector. These are exactly the properties of the frame 1-form (dreibein)
on a 3-manifold with tangent space group $SO(3)$ and metric
\beq\label{metric} G_{ij}\ =\ e^a_ie^a_j\ . \eeq
By straightforward computation one can show that
\bea\label{connection} \o^a_i\ &=&\ -{\d\O\over\d E^{ai}}\ =\
-{1\over2}\ \ee^{abc}\left\{ e^{bj}\pa_i e^c_j-e^{bj}e^c_k\G^k_{ij}
\right\}\nonumber\\*[8pt] &=&\ -{1\over2}\ \ee^{abc}\o^{bc}_i\ . \eea
Here $\G^k_{ij}$ is the Christoffel symbol for the metric $G_{ij}$, and
$\o^{ab}_i$ is just the standard spin connection on a Riemannian
3-manifold. Thus the composite gauge potential $\o^a_i$ of $SU(2)$ gauge
theory is the well-known spin connection, and we now see that a
conventional Riemannian spatial geometry underlies $SU(2)$ gauge theory.

A corollary of our discussion above is the fact that in three spatial
dimensions the spin connection is the variational derivative of the
local functional $\O [E]$ of Eq.\ (\ref{omegasu2}). This was established
in studies \cite{hns,nm} of the Ashtekar formalism for gravity in which
the form of $\O [E]$ with Eq.\ (\ref{littlee}) inserted was used, viz.,
\beq\label{3-form} \O [E]\ =\ {1\over2}\ \int d^3\!x\ \ee^{ijk}\
e^a_i(x)\pa_je^a_k(x)\ , \eeq
showing that $\O [E]$ is the integral of a natural 3-form.

Actually we have been a little too hasty in the above. The definition
Eq.\ (\ref{littlee}) actually implies that $\det E^{ai}\ge 0$, whereas
both signs of $\det E$ occur in gauge theory. So we should actually
define
\beq\label{littlee2} E^{ai}\ =\ \pm {1\over2}\ee^{ijk}\ee^{abc}
e^b_je^c_k\ , \eeq
with $\pm$ according to whether $\det E > 0$ or $< 0$. For each sign
above, there are two solutions for $e[E]$ which differ by a sign. We
make the convention to choose the solution with $\det e^a_i>0$, so that
we take
\beq\label{e} e^a_i\ =\ \pm\sqrt{|\det E^{ai}|}\ E^a_i \eeq
as the solution to Eq.\ (\ref{littlee2}). One can show that Eqs.\
(\ref{metric}) and (\ref{connection}) remain valid (but Eq.\
(\ref{3-form}) acquires a $\pm$ sign), so that $\o^a_i$ is the same
standard connection for both signs of $\det E$. Since $\o^a_i$ is an
even function of the frame, it can be reexpressed as an even function of
$E^{ai}$ and the sign in Eq.\ (\ref{e}) cancels.

Note that
\beq\label{densitized} E^{ai}\ =\ \pm e^{ai}\det e \eeq
is a ``densitized" inverse frame. One can show using Eqs.\
(\ref{connection}) and (\ref{littlee2}) that the total covariant
derivative vanishes, i.e.,
\beq\label{gradE} \nabla_iE^{ak}\ \equiv\
\pa_iE^{ak}+\G'^{k}_{ij}E^{aj}+\ee^{abc} \o^b_i E^c_k\ =\ 0\ , \eeq
where $\G'^k_{ij}$ is a not-often-used but standard connection for the
covariant differentiation of densities, namely
\bea\label{gammap} \G'^k_{ij}\ &=&\ -{1\over2}\d^k_j\pa_i\ln\det
G_{mn}+\G^{k}_{ij} \nonumber\\*[8pt] &=&\ {1\over4}(\d^k_i\pa_j
-\vf^{k\ell}\vf_{ij}\pa_l)\ln\det \vf^{mn}+{1\over2}\vf^{k\ell}
 \left[\pa_i\vf_{j\ell}+\pa_j
\vf_{i\ell}-\pa_\ell\vf_{ij}\right]\ , \eea
where, for reasons stated below, we have used the relation
$\vf^{ij}=\det G\ G^{ij}$ between the tensor density $\vf^{ij}$
introduced in the previous section and the inverse metric $G^{ij}$. One
can solve Eq.\ (\ref{gradE}) for $\o^a_i$ and obtain a form equivalent
to Eq.\ (\ref{connection}). The fact that $\nabla_iE^{ak}=0$ solidifies
the geometric interpretation of the electric field.

It is easy to see \cite{fhjl} that the curvature tensors of $\G '$ and
$\G$ co\"\i ncide, since the density term cancels:
\beq\label{riemann}
R^\ell_{\, kij}(\G ')\ =\ \pa_{[i}\G'^\ell_{j]k}+\G'^\ell_{m[i}
\G'^m_{j]k}\ =\ R^\ell_{\, kij}(\G )
\eeq
One can also show that the composite magnetic field, defined above
Eq.~(\ref{triv}) is related to the standard curvature by
\bea\label{Bhat} \hat{B}^{ai}\ &=&\ -{1\over2}\ \ee^{ijk}\ee^{abc}
R^{bc}_{jk}(\o )\nonumber\\*[6pt] &=&\ -{1\over2}\ {\det E \over \sqrt{
\det \vf}} \ee^{ijk}\hat
{\ee}_{mnq} E^{aq}R^{mn}_{\ \ \ jk}(\G ) \\*[8pt] &=&\ 2{\det E \over \sqrt{
\det \vf}}\ E^{aq}(R^i_q-{1\over2}\d^i_qR)\ . \eea
The standard curvature of the spin connection in the first line is
converted to space indices using the frame and (\ref{littlee2}), and the
representation of the curvature of a 3-manifold in terms of its Ricci and
scalar contractions
\beq\label{riemann3}
R_{ijk\ell}=G_{ik}R_{j\ell}- G_{i\ell}R_{jk}- G_{jk}R_{i\ell}
+ G_{j\ell}R_{ik}- {R\over2}( G_{ik}G_{j\ell}- G_{i\ell}G_{jk})
\eeq
is used in the final step. Note that $\hat{\ee}_{mnq}$ has components
$\pm 1,0$, and transforms as a tensor density of weight $-1$.

Let us now consider whether $E^{ai}$ or $e^a_i$, obtained through Eq.\
(\ref{e}), is the better variable for the dynamics of $SU(2)$ gauge
theory in this approach. Certainly $e^a_i$ is more geometric and has
lower dimension, but provisionally we prefer the electric field $E^{ai}$
because the parity transformation $E^{ai}(x)\to -E^{ai}(-x)$ is very
awkward to implement on $e^a_i$. So we shall use $E^{ai}, \vf^{ij}$ and
$\G'^k_{ij}$ for the rest of the paper. It is not difficult to convert
to $e^a_i, G_{ij}$ and $\G^{k}_{ij}$ if that proves to be desirable.

Finally we come to the question of implementing the Gauss law
constraint, $\bar{{\cal G}}^aF[E]=0$, within this approach to $SU(2)$
gauge theory. We shall describe several classes of gauge invariant
states, but we are not certain that they comprise the ``general
solution" of the constraint.

Following similar discussions \cite{fhjl,j} for the magnetic
representation, we note that $E^{ai}$ contains 9 components. Since
there are 3 gauge group ``angles", we would expect that it takes 6
functions to describe the gauge invariant content of an electric field
configuration. The symmetric tensor $\vf^{ij}$ has 6 independent
components. Although $\det E$ is another local gauge invariant, one has
$\det\vf=(\det E)^2$, and only the sign of $\det E$ is independent of
$\vf^{ij}$. However this sign is a complication for us. To handle it we
introduce $\vr(x)=\det E^{ai}(x)$ as an unconstrained field variable which
is a scalar density of weight 2. The most general functional of the local
invariants can then be written as $F[\vf^{ij},\vr]$ and the constraint
$\vr^2 =\det \vf$ will be enforced in the functional measure.

Let us consider the ``electric" Chern-Simons functional of the composite
spin connection
\beq\label{cs} C\!S[\o ]\ =\ {1\over16\p^2}\int d^3\!x\ \ee^{ijk}\left[
\o^a_i\pa_j \o^a_k+ {1\over3}\ \ee^{abc}\o^a_i\o^b_j\o^c_k\right]\ ,
\eeq
normalized to give $\hat{B}^{ai}=8\p^2\d (C\!S)/\d\o^a_i$. With $T\o$
denoting the finite gauge transformation of $\o$ under $E^{ai}\to
T^{ab}E^{bi}$, we have
\beq\label{csvar}
C\!S[T\o ]\ =\ C\!S[\o ]-{1\over96\p^2}
\int d^3\!x\ \ee^{ijk}\left[ T^{da}\pa_i T^{db} T^{eb}\pa_j T^{ec}
T^{fc}\pa_k T^{fa} \right]\ .
\eeq
The last term is the integer-valued winding number, so $C\!S[\o ]$ is
certainly infinitesimally gauge-invariant, and satisfies
$[{\cal G}^a(x),C\!S[\o ]\, ]=0$. But
\beq\label{csvar2}
C\!S[T_k\o ]\ =\ C\!S[\o ]+k
\eeq
for a gauge transformation $T_k$ with winding number $k$. All of the
above is standard \cite{rj}. One then sees that states of the form
\beq\label{thstates}
F[E,\th ]\ \equiv\ e^{i\th C\!S[\o ]}\ F[\vf ,\vr ]
\eeq
transform as
\beq\label{thtrsf}
F[T_kE,\th ]\ =\ e^{i k\th}\ F[E,\th ].
\eeq
Thus, as in the magnetic representation \cite{rj}, the Chern-Simons
functional, here a composite functional of $E^{ai}$, can be used to
relate states with nontrivial response to large gauge transformations to
invariant states, here $F[\vf ]$.

We also want to discuss briefly a third class of states which obey the
Gauss law constraint, namely functionals constructed from ``electric"
Wilson loops:
\bea\label{wilson} W[\o ,C]\ &=&\ {\rm Tr}\left[ {\rm P}\exp\ i\ \oint
dx^i\o_i\right] \nonumber\\*[8pt] \o_i\ &=&\ {1\over2}\t_a\o^a_i \eea
where $\t_1,\t_2,\t_3$ are Pauli matrices and $C$ is a closed curve in
$\IR^3$. Certainly $[{\cal G}^a(x),W[\o ,C]\, ]=0$, and state
functionals formed from $W[\o ,C]$ satisfy the gauge constraint.
Nevertheless, it appears that these states are already included in the
class $F[\vf]$. The reason for this is that from the relation
$\vf^{ij}(x)=E^{ai}(x)E^{aj}(x)$, the tensor field $\vf^{ij}(x)$ along a
curve $C$ determines the electric field up to multiplication by an
$O(3)$ matrix $R^{ab}(x)$. Since $\o^a_i$ is quadratic in $E^{ai}$, it
is not sensitive to an improper factor in $R^{ab}$, so $\o^a_i$ is
determined up to an $SO(3)$ gauge transformation, and $W[\o,C]$ is thus
uniquely determined by $\vf^{ij}$. Independently of this one can ask the
general question of the relation between electric
Chern-Simons and Wilson loop functionals and their magnetic analogues.
They do not appear to be simply related by the functional Fourier
transform \cite{gj} between magnetic and electric representations of the
theory.

We now wish to discuss the form of the Hamiltonian
$\bar{H}$ of Eq.~(\ref{eq Hbar}) acting on states $F[\vf ,\vr ]$. We need to
express $\bar{B}^{ai}(x)F[\vf ,\vr ]$ of Eq.~(\ref{eq Bbar}) in terms of
$\vf^{ij}$ and $\vr$ using the chain rule
using the chain rule
\bea\label{chainrule}
{\d\over\d E^{ak}}F[\vf ]\ &=&\  {\d\vf^{pq}\over\d E^{ak}}
{\d F\over\d\vf^{pq}} + {\d\vr\over\d E^{ak}}
{\d F\over\d \vr}\nonumber\\*[8pt]
&=&\ 2E^{ap}{\d F\over\d\vf^{pq}} + \vr E^a_k {\d F\over\d\vr}
\eea
Using also Eqs.~(\ref{gradE}-\ref{Bhat}) one can obtain by a straightforward
calculation the expression
\bea\label{BF}
\bar{B}^{ai}F[\vf ,\vr ]\ &=&\ 2\left\{\, {1\over g} {\vr \over \sqrt{ \det
\vf}} \,
E^{ap}(R^i_p- {1\over2}\d^i_p R)+
i\ee^{ijk}\left( E^{ap}\nabla_j{\d\over\d\vf^{kp}}+{\vr \over 2} E^a_k
\nabla_j {\d \over\d \vr}\right)\right.\cr
&& - g\ee^{ijk}\ee^{pqr} \vr E^a_r \ \left(
{\d^2\over\d\vf^{jp}\d\vf^{kq}} + \vr \vf^{kq} {\d^2\over\d\vf^{jp}\d \vr}
\right) \cr && \left. -{1 \over 2}g \vr \ E^{ai} {\d^2 \over \d \vr \d \vr}
- g \ E^{ai} \d (0) {\d \over \d \vr}\right\}F[\vf , \vr ]
\eea
We have dropped the $\d (0)$ ordering term of Eq.~(\ref{eq Bbar}) in Eq.\
(\ref{BF}), because of the provisional conclusion of the Appendix, that this
term vanishes after regularization.
The salient feature of this equation is that spatial covariant derivative
appropriate to the tensor and density character of $\d F\ / \ \d\vf^{kp}$ and
$\d F \ / \ \d \vr$ have automatically appeared through (\ref{gradE}). These
derivatives are
\bea\label{nablaF}
\nabla_j{\d F\over\d\vf^{kp}}\ & = & \ \pa_j{\d F\over\d\vf^{kp}}-
\G'^q_{jp}{\d F\over\d\vf^{kp}}-\G^{q}_{jk}{\d F\over\d\vf^{pq}} \nonumber
\\*[8pt] \nabla_j{\d F \over \d \vr}\ & = & \ (\pa_j +{1 \over 4} \pa_j
\ln \det \vf){\d F \over \d \vr}
\eea
A new $\d (0)$ term has also appeared in (\ref{BF}). Since its coefficient
is covariant, it does not vanish by the point-spitting argument used in the
appendix. We can say little more about this term now except that it should
be studied in the context of a more systematic regularization procedure.

We now consider the magnetic energy density
\beq\label{energy}
{\cal E}_M\ F[\vf ]\ =\ {1\over2} \d_{\bar{\imath}i}\bar{B}^{a\bar{
\imath}}\bar{B}^{ai} F[\vf ]\ .
\eeq
Even without a detailed computation, one sees that the gauge indices
cancel, e.g. $E^{ai}E^{aj}=\vf^{ij}$, so that the full Hamiltonian can
be rewritten entirely in terms of the spatial geometric variables $\vf
,\G '$ and $R$ together with $\vr$ which is also essentially geometric.
The elimination of all non gauge-invariant variables is is a remarkable
transformation of the original theory, although the result is complicated
and the geometrization imperfect because of the presence of $\vr$.

The Hamiltonian simplifies a great deal if one restricts to wave functions
$F[\vf ]$ which are independent of $\vr$. Not only do 4 of the 7 terms in
(\ref{BF}) drop, but the remaining $\vr$ dependence is actually helpful. Namely
all imaginary ``interference terms'' in ${\cal E}_M$ cancel in the sum of
the two configurations $\vr (x)$ and $-\vr (x)$, and ${\cal E}_M$ becomes a
sum of two positive terms depending only on $\vf$. Although the restriction
to $F[\vf ]$ does not follow from any symmetry, it could possibly be
justified a posteriori in the vacuum sector because it leads to a lower
variational energy than $F[\vf ,\vr ]$. In any case the simplifying
assumption may be useful in a first exploration of the dynamics.

One can also transform the functional measure used to compute matrix
elements of $\bar{H}$ in states $F[\vf ]$. Dimensional and $GL(3)$
symmetry arguments are sufficient to give at each point $x$
\bea\label{measure}
\prod_{a,k} dE^{ak}(x)\ &=&\ d \vr (x) \ \prod_{i \le j } d \vf^{ij}(x)
\d(\vr(x)^2- \det \vf(x))
\eea
This is to be understood as an identity valid when the integrand depends only
on $\vf$ and $\vr$. An irrelevant numerical constant has been dropped.

The phase $\O [E]$ of Eq.~(\ref{omegasu2}) involves the matrix inverse
of the electric field, so our transformation is singular when $\det
E^{ai}=0$.  The composite connection $\o^a_i$ as well as $\G '^k_{ij}$
are also singular here.  One can see upon closer inspection of the
magnetic energy density Eq.~(\ref{BF}) that the singular terms always
involve spatial derivatives $\pa_i\vf^{jk}$.  As in \cite{gj,j} we
believe that these singularities are the functional analogue of the
angular momentum barrier for central forces in quantum mechanics.  Any
finite energy wave functional must ``know how to behave itself" as such
singular field configurations are approached, otherwise it would not
have finite energy.\bigskip

\section{general gauge groups}\bigskip

The extension of the present methodology to gauge groups larger than
$SU(2)$ is important for two reasons. First the realistic color group of
the strong interactions is $SU(3)$. Second, we must show that the
geometrization found for $SU(2)$ is not an accidental consequence of the
fact that $SO(3)$ ($\approx SU(2)$) is the tangent space group of a
3-dimensional Riemannian space.

Technically, it was easy to construct the phase $\O [E]$ for $SU(2)$
because the electric field $E^{ai}(x)$ is a $3\times 3$ matrix with a
matrix inverse $E^a_i(x)$ which respects gauge and $GL(3)$ covariance.
For larger groups, $E^{ai}(x)$ is a rectangular matrix, and there is no
inverse. The major problem in constructing the phase $\O [E]$ for other
semi-simple groups is to find an appropriate substitute for the inverse.
In this section we present such a construction.

To begin with, we
attempt to generalize Eq.\ (\ref{omegasu2}) where, however, since we do
not have an $E^a_i(x)$ available, we write instead
\beq\label{omegasu3}
\O [E]\ \equiv\ {1\over2}\ \int d^3\! x\ {f^{abc} E^{ai}(x)
E^{bj}(x)\over (\det \vf )^{1/4}}\ \pa_iR^c_j(x)
\ \equiv\ {1\over 2}\int d^3\! x\ \ee^{ijk}L^a_i(x)\pa_jR^a_k(x)\ ,
\eeq
with the variable $R^a_i(x)$ to be determined so that $\O [E]$ is
$GL(3)$ invariant with gauge variation Eq.\ (\ref{eq varo}). The
quantity $L^a_i(x)$ above is simply shorthand for
\beq\label{L}
L^a_i(x)\ =\ {1\over2}\ \hat{\ee}_{ijk}\ {f^{abc} E^{bj}(x)
E^{ck}(x)\over (\det \vf )^{1/4}}\ .
\eeq
We have divided by $(\det \vf )^{1/4}$ in order to make $L^a_i(x)$ a
covariant vector rather than a density, and we see from the last
equality in Eq.\ (\ref{omegasu3}) that $\O [E]$ is the integral of a
3-form, and therefore $GL(3)$ invariant, if $R^a_i(x)$ is also a
covariant vector. Note that it was not necessary to insert the
determinantal factor for $SU(2)$ because $R^a_i(x)$ in that case is the
matrix inverse of $E^{ai}(x)$, and this was sufficient for $GL(3)$
invariance. $R^a_i(x)$ is now fixed as a function of $E^{ai}(x)$ by our
first requirement on $\O [E]$, namely, that it satisfy Eq.\ (\ref{eq
varo}). We now examine that requirement. The gauge variation of $\O
[E]$ in Eq.\ (\ref{omegasu3}) is easily computed if we assume that
$R^a_i(x)$ transforms in the adjoint representation:
\beq\label{omegavar}
\d\O [E]\ =-{1\over 2}
\int d^3\! x\ \ee^{ijk}\ f^{abc}\ L^a_i(x)R^b_j(x)\pa_k\th^c(x)\ .
\eeq
The requirement that this is of the form of Eq.\ (\ref{eq varo}) gives
the following condition on $R^a_i(x)$:
\beq\label{MR=E}
{1\over 2}\ee^{ijk}\ f^{abc}\ L^a_i(x)R^b_j(x)\ \equiv\
M^{ck,bj}(x)R^b_j(x)\ =\ E^{ck}(x)\ .
\eeq
This is a linear system and there is a unique solution for $R^a_i(x)$
provided that the determinant of the $3\dim G\times 3\dim G$ direct
product matrix $M$ is non-vanishing. It is also easy to show from the
structure $R=M^{-1}E$ that $R^a_i(x)$ has the required
gauge and $GL(3)$ properties assumed above. An analytic calculation of
$M^{-1}$ would be necessary to have a truly explicit construction of the
phase $\O [E]$. This is a difficult task, and we shall be content here
with the fact that we have reduced the problem to this point.

We end this section with a possible alternative procedure to determine
the phase $\O [E]$. Again, faced with the same initial problem of not
having an ``inverse" electric field $E^a_i$, we try another
generalization of the $SU(2)$ phase, by writing an {\it ansatz}
identical in form to Eq.~(\ref{3-form}):
\beq\label{3formsu3}
\O [E]\ =\ {1\over2}\int d^3\!x\ \ee^{ijk}\ e^a_i(x)\pa_je^a_k(x)\ ,
\eeq
with the difference that now the variables $e^a_i(x)$ form a
$3\times\dim G$
matrix, as yet undefined. The requirement that this phase has the
correct gauge transformation Eq.~(\ref{eq varo}), then determines
$e^a_i$ implicitly in a similar way as for $R^a_i$ above. This
requirement reads:
\beq\label{E=ee}
E^{ai}={1\over2}\ee^{ijk}f^{abc}e^b_je^c_k\ .
\eeq
One must then solve this set of $3\dim G$ quadratic equations to obtain
$e[E]$.  We have not been able to do this (despite considerable effort
for the group $SU(3)$), but we find that it is an intriguing algebra
problem with a group-theoretic flavor.  It is formally identical to the
problem of finding, for a general group $G$, the gauge potential $A^a_i$
($e^a_i$ here) given a constant magnetic field $B^{ai}$ (here $E^{ai}$).
The solution to this would yield a phase $\O [E]$ which would
automatically have the proper gauge and $GL(3)$ transformation
properties, and could possibly lead to a simpler formulation of the
theory than the one based on Eq.~(\ref{omegasu3}).\bigskip

\section{$SU(3)$ gauge theory}\bigskip

We now explore the $SU(3)$ theory in order to ascertain the spatial
geometry associated with a larger gauge group. The first step is to use
the group theory and the physics to define a basis of eight vectors for
the adjoint representation of the group. The basis is then used to
define the connection, torsion, and curvature of the geometry. Then we
identify the class of gauge invariant states analogous to $F[\vf^{ij}]$
of Sec.~3, and show that the Hamiltonian acting on these states can be
expressed in terms of gauge invariant and geometric quantities. The
attitude we shall take is that all geometric information is
contained in the $SU(3)$ gauge connection $\o^a_i$ calculated from $\O
[E]$ in Eq.~(\ref{omegasu3}). The basis of eight vectors is a
generalized frame used to transfer this information to geometric
variables with spatial indices only. This attitude is consistent
with the situation for $SU(2)$, but little thought was required there
because the geometry was completely standard.

The first three 8-vectors of the basis are simply the three spatial
components $E^{ai}$ of the electric field. These are linearly
independent for generic field configurations in which the rectangular
matrix has rank 3. Using the $d$-symbols of $SU(3)$ we
construct six additional 8-vectors
\beq\label{eextended}
E^{ajk}\ \equiv\ d^{abc} E^{bj}E^{ck}\ .
\eeq
First, we orthogonalize these with respect to the first three by
defining
\beq\label{hatE}
\hat{E}^{ajk}\ \equiv\ E^{ajk}-E^{am}\vf_{mn}\vf^{njk}\ ,
\eeq
where $\vf_{mn}$ is the matrix inverse of $\vf^{ij}$ and
\beq\label{phi3}
\vf^{ijk}\ \equiv\ d^{abc}E^{ai}E^{bj}E^{ck}\ .
\eeq
The six $\hat{E}^{ajk}$ span an orthogonal subspace to that of
$E^{ai}$. Within that subspace, the trace $\hat{E}^a=\hat{E}^{amn}
\vf_{mn}$ is generically linearly related to the 5 traceless combinations
\beq\label{eexttrless}
\hat{E}^{a\{ ij\} }\ \equiv\ \hat{E}^{aij}-{1\over
3}\vf^{ij} \hat{E}^a\ ,\eeq
and these are generically linearly independent. So as a basis of 8
vectors we take the set
\beq\label{basis}
\{\ E^{ai},\hat{E}^{a\{ jk\} }\ \}\ . \eeq
when the mutual orthogonality is useful, and otherwise the set
\beq\label{basis2}  \{\, E^{ai}\, ,  E^{a\{ ij\}}=
E^{aij}-{1\over3}\vf^{ij} \vf_{mn} E^{amn}\, \}. \eeq
We shall not characterize precisely the non-generic configurations in
which the five $E^{a\{ jk\}}$ fail to be linearly independent. Presumably
this occurs when the span of any two of the three vectors $E^{ai}$
determines an $SU(2)$ subalgebra of $SU(3)$.

Connections for the $SU(3)$ geometry are defined by the pair of equations
\bea\label{DE1}
\hat{D}_iE^{ak}\ &\equiv&\ -\hat{\G}'^k_{ij}E^{aj}-T_i^{ak} \\
\label{DE2} E^{aj}T_i^{ak} &\equiv& 0\ ,\eea
which is equivalent to the fact that $\hat{D}_iE^{ak}$ can be expanded
uniquely
in the basis of Eq.~(\ref{basis}). Note that Eqs.~(\ref{DE1}-\ref{DE2})
comprise $72+27$ equations for $27+72$ components of $\hat{\G}'$ and
$T$. One can see that $\hat{\G}'^k_{ij}$ transforms as a connection (for
densities of weight one), while $T_i^{ak}$ is a gauge adjoint $GL(3)$
tensor density.

We now contract Eq.~(\ref{DE1}) with $E^{a\ell}$ and symmetrize in
$k\ell$, which leads to
\beq\label{metrcomp}
\pa_i\vf^{k\ell}+\hat{\G}'^k_{ij}\vf^{j\ell}+\hat{\G}'^\ell_{ij}
\vf^{kj}=0\ .
\eeq
This is simply the metric compatible relation between $\hat{\G}'$ and
the ``densitized metric" $\vf_{ij}$. It then follows from simple algebra
that $\hat{\G}'$ takes the form of a (densitized) connection with
torsion, namely
\beq\label{connectionsu3}
\hat{\G}'^k_{ij}=\G '^k_{ij}-K^k_{ij}\ ,
\eeq
where $\G '^k_{ij}$ is just the Riemannian $\G '$ of Eq.~(\ref{gammap})
and $K$ is the contortion tensor, which satisfies the antisymmetry
property
\bea\label{antisymm}
K_{ijk}&=&-K_{ikj}\\
K_{ijk}&\equiv&K_{ij}^{\ \ \ell} G_{\ell k}\ .
\eea

Because of orthogonality to $E^{ai}$, $T_i^{ak}$ is determined entirely
by the spatial tensor density
\beq\label{KT}
K_i^{\ \{ mn\} k}\equiv \hat{E}^{a\{ mn\}}T_i^{ak}=-\hat{E}^{a\{
mn\}}\hat{D}_iE^{ak}\
. \eeq
Using the components $\hat{E}^a_i,\hat{E}^a_{\{ mn\}}$ of the $8\times8$
matrix inverse of the basis Eq.~(\ref{basis}) we see that
\beq\label{T}
T_i^{ak}=\hat{E}^a_{\{ mn\}}K_i^{\ \{ mn\} k}\ .
\eeq
We regard $T_i^{ak}$ or $K_i^{\ \{ mn\} k}$ as a new type of torsion.
The Bianchi identity Eq.~(\ref{Bian}) implies that all torsions are
traceless:
\beq\label{traceless}
K_{ij}^{\ \ i}=T_i^{ai}=K^{\{ mn\} i}_i=0\ .
\eeq

The torsions are local functions of $E$ and $\pa E$ which can be found
from the definition Eq.~(\ref{DE1}-\ref{DE2}), once we have the explicit
form of $\o^a_i$. In turn this requires the construction of the matrix
$R^a_i[E]$ which enters the phase $\O [E]$ of Eq.~(\ref{omegasu3}). Note
that Eq.~(\ref{DE1}) can be expressed in terms of the total covariant
derivative $\nabla_i$ of Eq.~(\ref{gradE}), but now we have
\beq\label{covE}
\nabla_iE^{ak}=-T_i^{ak}\, ,
\eeq
so the frame is no longer covariantly constant, but
$\vf_{ij}$ and also $G_{ij}$ are, since Eq.~(\ref{metrcomp})
is equivalent to
\beq\label{covf}
\nabla_i\vf^{jk}=0\ .
\eeq

The next step is to study the curvature by taking a further gauge
derivative of Eq.~(\ref{DE1}) and antisymmetrizing to obtain
\bea\label{curvsu3}
[\hat{D}_i,\hat{D}_j]E^{ak}&=&-\left\{ R^k_{\ \ell
ij}E^{a\ell}+\hat{D}_{[i}T_{j]}^{ak}
+\G '^k_{[i\ell} T_{j]}^{a\ell}\right\}\nonumber\\*[8pt]
&=&f^{abc}\hat{\ee}_{ijm}\hat{B}^{bm}E^{ck}\ .
\eea
The sum of the last two terms in the first line is $GL(3)$ covariant,
and we have used the gauge Ricci identity to obtain the last line.

We now wish to obtain the $SU(3)$ generalization of Eq.~(\ref{Bhat}) and
express the composite magnetic field $\hat{B}$ in terms of the curvature
and torsion. This is awkward because $E^{ck}$ itself does not have an
inverse, but the full frame Eq.~(\ref{basis2}) can be brought to use as
follows. The gauge covariant derivative of Eq.~(\ref{eextended}) can be
evaluated as
\beq\label{dereext}
\hat{D}_jE^{ak\ell}=-\G '^k_{jm}E^{am\ell}-\G
'^\ell_{jm}E^{akm}+2d^{abc}E^{bk}T_{j}^{c\ell}
\eeq
and one also finds
\bea\label{curvsu3ext}
[\hat{D}_i,\hat{D}_j]E^{ak\ell}&=&-\left\{ R^k_{\
mij}E^{am\ell}+R^\ell_{\
mij} E^{akm}+2d^{abc}E^{bk}\left(\hat{D}_{[i}T_{j]}^{c\ell}+
\G '^\ell_{[im} T_{j]}^{cm}\right)\right\}\nonumber\\*[8pt]
&=&f^{abc}\hat{\ee}_{ijm}\hat{B}^{bm}E^{ck\ell}\ .
\eea
With $\{...\}$ denoting symmetrization and removal of the trace, we
obtain
\beq\label{BEext}
f^{abc}\hat{\ee}_{ijm}\hat{B}^{bm}E^{c\{ k\ell\}}=
-\left\{ R^{\{ k}_{\ mij}E^{a\ell\} m}+
d^{abc}E^{b\{ k}\left(\hat{D}_{[i}T_{j]}^{c\ell\} }+
\G '^{\ell\} }_{[im}T_{j]}^{cm}\right)\right\}\ .
\eeq
Using the components of the inverse matrix $E^a_k,E^a_{\{k\ell \}}$
it is now simple to obtain $\hat{B}$ from Eqs.~(\ref{curvsu3},
\ref{BEext}):
\bea\label{Bsu3}
\hat{B}^{ai}&=&-{1\over6}f^{abc}\ee^{ijk}\left\{
E^b_m\left[
R^m_{\ \ njk}E^{cn}+\hat{D}_{[j}T_{k]}^{cm}+
\G '^m_{[jn} T_{k]}^{cn}\right]\right.\nonumber\\*[8pt]
& &\left.+E^b_{\{ mn\}}\left[
R^m_{\ \ell ij}E^{cn\ell}+d^{cde}E^{bm}\left(
\hat{D}_{[j}T_{k]}^{cn}+
\G '^n_{[j\ell} T_{k]}^{c\ell}\right)\right]\right\}\ .
\eea
This is the desired expression for the composite magnetic field. One
can go further and substitute the representation of Eq.~(\ref{riemann3})
for $R^m_{\ \ njk}$, which holds with torsion \cite{fhjl}, and one can
use Eq.~(\ref{T}) to express $DT+\G'T$ in terms of the total spatial
covariant derivative of $K_i^{\ \{ mn\} k}$ and $\nabla \hat{E}^a_{\{
mn\}}$.
We shall not write the final resulting formula. Note that the matrix $M$
of Eq.~(\ref{MR=E}) is singular for electric fields which vanish except
in an $SU(2)$ subalgebra of $SU(3)$, and Eq.~(\ref{Bsu3}) is also
singular in this case.

The next stage of the discussion concerns gauge invariant states and
local variables \cite{j} for $SU(3)$.  We start with the observation
that the gauge invariant content of an $SU(3)$ electric field
configuration can be described by $24-8=16$ variables.  The symmetric
tensor densities $\vf^{ij}$ and $\vf^{ijk}$ contain precisely $6+10=16$
algebraically independent components. Hence all invariants are
algebraic functions of these fundamental ones. Among the other local
invariants which appear in the Hamiltonian are the ``efterminant"
and ``extended metric"
\bea\label{eft}
{\rm eft}\ E &\equiv& {1\over6}f^{abc}\hat{\ee}_{ijk}E^{ai}E^{bj} E^{ck}
\nonumber\\
\vf^{jk;\ell m} &\equiv& E^{a jk}E^{a\ell m}\ .
\eea
However both practically and theoretically the study of relations among
$SU(3)$ invariants is more complicated than in the $SU(2)$ case. For instance,
the ring of local polynomial invariants is not a finite extension of the
ring of polynomials in $\vf^{ij}$ and $\vf^{ijk}$, although this problem can
be circumvented. Moreover only a finite number of invariants can appear in
the Hamiltonian and the full complexity of the ring of invariants is
probably not needed. To give a more definite answer to these questions, we
would need an explicit expression for the $SU(3)$ phase.

In the following,
we are nevertheless able to illustrate partly our purpose (geometrization)
by working out the expectation value of the Hamiltonian between states
of the form $F[\vf^{ij},\vf^{ijk}]$ (generalizations of the states
$F[\vf^{ij}]$ considered for $SU(2)$) depending only on the simplest
fundamental invariants. Even if they are not the most general states, they
might encode some interesting physics. We need the chain rule
\bea\label{chainsu3}
{\d\over\d E^{ak}}F\ &=&\
{\d\vf^{pq}\over\d E^{ak}}{\d F\over\d\vf^{pq}}+
{\d\vf^{pqr}\over\d E^{ak}}{\d F\over\d\vf^{pqr}}\nonumber\\*[8pt]
&=&\ 2\, E^{ap}{\d F\over\d\vf^{pq}}+
3\, d^{abc}E^{bp}E^{cq}
{\d F\over\d\vf^{pqk}}\ .
\eea
After some algebra one finds that the second term in $\bar{B}^{ai}F$ of
Eq.~(\ref{eq Bbar}) can be expressed in terms of $SU(3)$ connections and
torsions as
\bea\label{BbarF}
i\ee^{ijk}\hat{D}_j{\d F\over\d E^{ak}}&=&i\ee^{ijk}\left\{
2E^{ap}\nabla_j{\d F\over\d\vf^{pk}}-T_j^{ap}{\d F\over\d\vf^{pq}}
\right.\nonumber\\*[8pt]
& &\left. +3\, E^{apq}\nabla_j{\d F\over\d\vf^{pqk}}-
6\, d^{abc}E^{bp}E^{c}_{mn}K_j^{\{ mn\} q}{\d F\over\d\vf^{pqk}}\
\right\}\ ,\eea
where $\nabla_j{\d F\over\d\vf^{pk}}$ has been defined in
Eq.~(\ref{nablaF}), and
\beq\label{covFsu3}
\ee^{ijk}\nabla_j{\d F\over\d\vf^{pqk}}=\ee^{ijk}\left[
\pa_j{\d F\over\d\vf^{pqk}}-\G'^r_{jp}{\d F\over\d\vf^{rqk}}
-\G'^r_{jq}{\d F\over\d\vf^{prk}}\right]\ .
\eeq
Similarly, the third term in Eq.~(\ref{eq Bbar}) can be written as
\bea\label{dFdE2}
-{g\over2}\ee^{ijk}f^{abc}{\d^2F\over\d E^{bj}\d E^{ck}}&=&
-{g\over2}\ee^{ijk}f^{abc}\left\{
4E^{bp}E^{cq}{\d^2F\over\d \vf^{pj}\d\vf^{qk}}\right.\nonumber\\*[8pt]
&&\left.+12E^{bp}E^{crs}{\d^2F\over\d \vf^{pj}\d\vf^{rsk}}+9
E^{bpq}E^{crs}{\d^2F\over\d \vf^{pqj}\d\vf^{rsk}}\right\}\ .
\eea
No $\d (0)$ ordering terms arise in Eq.~(\ref{dFdE2}), and we assume
that the fourth term in Eq.~(\ref{eq Bbar}) vanishes after regularization
as discussed in the Appendix for $SU(2)$.

Consider now the magnetic energy density Eq.~(\ref{energy}) with each
factor $\bar{B}^{ai}$ expressed as the sum of
Eqs.~(\ref{Bsu3}-\ref{BbarF}-\ref{dFdE2}).  It is clear that all gauge
indices are contracted out in local invariant variables such as
$\vf^{pq;rs}$, $f^{abc}E^{apq}E^{br} E^{cs}$ and several others. So the
expectation value involves the fundamental invariants and possibly some
auxiliary ones which may (or may not) disappear (as in the SU(2) case) if
one sums over all configurations $E^{ai}$ giving the same value of $\vf^{ij}$
and $\vf^{ijk}$. This discussion also applies to the torsions $K_{ij}^{\ \ k}$
and $K_i^{\{ jk\} \ell}$ which should be expressible in terms of the
fundamental invariants, some auxiliary ones, and their first spatial
derivatives (in torsion-free covariant combinations). Symbolic manipulation
programs can be useful to help find the required expressions which are
necessary to express the $SU(3)$ gauge theory in complete geometric form.

This discussion has shown that our geometric ideas can be extended to
the gauge group $SU(3)$, and that there is an interesting spatial
geometry associated with this realistic color group.  The theory is not
yet in entirely explicit form.  For this one must obtain the matrix
$R^a_j$ and the inverse frame components for Eqs.~(\ref{basis}) and
(\ref{basis2}) as functions of $E^{ai}$, and one must solve the problem
of independent $SU(3)$ invariants discussed in the previous paragraph.
These ``mechanical'' problems are not necessarily easy, and we believe
that the effort to solve them is justified only if the spatial geometry
is shown to be useful for the dynamics in the $SU(2)$ theory of Sec.~3,
which is far simpler.\bigskip

\section{Discussion}\bigskip

We have shown that it is possible to reexpress the geometry of
non-abelian gauge theories in terms of a 3-dimensional spatial geometry.
The first and most important step was the unitary transformation $\Ps
[E] =\exp (i\O [E]/g) F[E]$ which allowed us to impose the Gauss law
constraint on $F[E]$ and to exploit the fact that $\o ^a_i=-\d \O / \d
E^{ai}$ transforms as a composite gauge connection.

For gauge group $SU(2)$, $\o ^a_i$ is just the standard spin connection
of a Riemannian 3-manifold.  We were naturally led to define metric- and
connection-like variables $\vf ^{ij}$ and $\G'^i_{jk}$ which are
equivalent to the ordinary Riemannian metric and Christoffel connection.
The $SU(2)$ theory essentially geometrizes itself, and a conventional
Riemannian geometry underlies the theory.

For larger gauge groups, and for $SU(3)$ in particular, the same
approach leads to a metric-preserving geometry with torsion of both
standard and novel type.  The construction of Secs.~4 and 5 was not
quite explicit because certain ``mechanical problems'' of analytic
matrix inversion and relations among group invariants remain to be
solved.  Apart from these problems, it is also possible that another
choice of phase $\O [E]$ or basis $E^{ai},\; E^{a\{ ij\}}$ could lead to
a simpler formulation.

Our initial motivation, beginning in \cite{fhjl}, was to express the
Hamiltonian in gauge invariant variables in order to develop a new
approach to the non-perturbative dynamics of gauge theories.  What has
been achieved so far is just a formal structure, of some elegance we
believe, but there are many difficulties to be overcome before it can be
applied to real physics.  The non-linear transformation to variables
$\vf ^{ij}=E^{ai}E^{aj}$ may exacerbate the problem of Lorentz
covariance in the Hamiltonian formalism.  A suitable cutoff procedure
must be found and one must cope with a Hamiltonian which is up to fourth
order in functional derivatives.  The fundamental unitary transformation
is non-perturbative, so the composite magnetic field $\hat{B}^{ai}$
appears in (\ref{eq Bbar}) with coefficient $1/g$, and there are
singular terms up to order $1/g^2$ in the Hamiltonian, as in
\cite{gj,iksf}.  These terms make it problematic to perform short
distance calculations to test whether the transformed theory has the
expected short distance behavior.  But since these singular terms are
the result of the exact treatment of the non-abelian gauge invariance,
they may represent a significant non-perturbative aspect of the theory.
Finally, the notion \cite{gj} that the behavior of physical wave
functions at the singular points of the unitary transformation used is
controlled by the energy barrier terms in H requires exploration.  All
of these problems appear to be substantial but we hope that the
geometric structure of the formal theory provides the impetus to solve
them.\bigskip

\section*{acknowledgments}
D.Z.F. and P.H. warmly thank their collaborators on Ref. \cite{fhjl}
K. Johnson and J.-I. Latorre for extensive discussions from which several
of the ideas used here emerged. They also thank S. Forte and E. Moreno
for useful discussions concerning the Chern-Simons functional.

\appendix

\section{}\bigskip

In this appendix we show explicitly for gauge group $SU(2)$ that the
singular term in $\bar{B}^{ai}$, when properly regularized, vanishes.
The singular term comes from the ill-defined quantity
\beq \ee ^{ijk}\ee ^{abc} \frac{\d \o^c_k(x)}{\d E^{bj}(x)} \eeq
It is easy to see that the rest of $\bar{B}^{ai}$ has the gauge and
tensorial properties of a magnetic field. Formally the singular term
also does. So we have to look for a regularization that preserves
these properties. The most obvious candidate would be to point-split,
i.e., work with
\beq \ee ^{ijk}\ee ^{abc} \frac{\d \o^c_k(y)}{\d E^{bj}(x)} \eeq
and take the limit $x=y$. However the quantity
\beq \label{eq posp} \frac{\d \o^c_k(y)}{\d E^{bj}(x)} =-\frac{\d
^2 \O}{\d E^{bj}(x) \d E^{ck}(y)}\eeq
does not transform as a geometric object at point $x$ but as a
``bi-geometric'' object at points $x$ and $y$ (a gauge and
contravariant spatial vector at $x$ and $y$). This is clear from its
definition but can also be checked on the explicit form of the second
variation of $\O$ involving $\d (x-y)$ and its first derivative. So
the contraction of (\ref{eq posp}) with $\ee^{ijk} \ee^{abc}$, which
is covariant with respect to gauge and $GL(3)$ transformations at a
single point, is not geometric. This is significant because (\ref{eq
posp}) is singular as $y \leftarrow x$.

A remedy for this is to introduce a linear operator $M^{k'c c'}_k(x,y)$
such that if $T^{c'}_{k'}$ is a gauge and contravariant spatial vector
at $y$ then $M^{k'c c'}_k(x,y)T^{c'}_{k'}=\tilde{T}^c_k$ has the same
geometric properties at $x$. Then
\beq M^{k'c c'}_k(x,y)\frac{\d ^2 \O}{\d E^{bj}(x) \d E^{c'k'}(y)}
\equiv \left[\frac{\d ^2 \O}{\d E^{bj}(x) \d E^{ck}(y)} \right]^{(cov)}
\eeq
will be a geometric object at $x$.

In general a smooth choice of $M$ is possible only locally. One must
choose gauge and affine connections, and use these to
parallel-transport $T^{c'}_{k'}$ along a path from $y$ to $x$. So
there are many ambiguities in the definition of $M$. But as stressed
above,
\beq \frac{\d ^2 \O}{\d E^{bj}(x) \d E^{ck}(y)} \eeq
is a local distribution of order $1$, so that all what is needed is
$M^{k'cc'}_k(x,x)$ and $\left(\pa^{(y)} M^{k'c c'}_k\right)(x,x)$,
and this only involves the gauge and affine connections at point $x$.

To compute the second variation of $\O$, the simplest way is to Taylor
expand $\O [E+E'+ E'']$ to first order in $E'$ and $E''$. The result
is
\bea \label{eq 2varo} \lefteqn{\int \int d^3\!x\ d^3\!y\ \frac{\d ^2 \O}
{\d E^{bj}(x) \d E^{ck}(y)} E'^{bj}(x) E''^{ck}(y)=} \nonumber \\*[6pt]
 & & -\frac{1}{2} \int d^3\!z\ \ee^{def} \ba[t]{l} \left[ E'^{d\ell}
(\pa_{\ell}E''^{em}) E^f_m - E^{d\ell} (\pa_{\ell} E''^{em}) E^g_m
E'^{gn} E^f_n -E'^{d\ell}(\pa_{\ell} E^{em}) E^g_m E''^{gn} E^f_n \right.
\\*[6pt] \left.+ E^{d\ell}(\pa_{\ell} E^{em}) E^h_m E'^{hn} E^g_n E''^{gp}
E^f_p + 4\mbox{ terms with } E' \leftrightarrow E'' \right]\ . \ea \eea
Now, because the left-hand side is a geometric object, the right-hand
side does not change if one replaces everywhere ordinary partial
derivatives by total (gauge and affine) covariant derivatives acting
on densities, making every term geometric.

As we have seen in Sec. 3, the electric formulation of the
$SU(2)$ theory has brought natural (gauge and affine) connections to
the fore, and it is more than natural to use these to define $M$ and
to rewrite (\ref{eq 2varo}). In the covariant form
\bea \label{eq varia} \lefteqn{\int \int d^3\!x\ d^3\!y\
\frac{\d ^2 \O}{\d E^{bj}(x) \d E^{ck}(y)} E'^{bj}(x) E''^{ck}(y)=}
\nonumber\\*[8pt]
& & -\frac{1}{2} \int d^3\!z\ \ee^{def} \ba[t]{l} \left[
E'^{d\ell}(\pa_{\ell} E''^{em}+ \G'^m_{\ell p}E''^{ep}+
\ee^{eah}\o^{a}_{\ell}
E''^{hm}) E^f_m
\right.
\\*[8pt] \left.
- E^{d\ell}(\pa_{\ell} E''^{em}+ \G'^m_{\ell p}E''^{ep}+
\ee^{eah}\o^{a}_{\ell} E''^{hm}) E^g_m E'^{gn} E^f_n
+2\mbox{ terms with } E' \leftrightarrow E'' \right]\ . \ea \eea

The second derivative of $\O$ is obtained by substituting $\d^{ab}
\d^i_j \d(x-z)$ for $E'^{ai}(z)$ (resp. $\d^{ac} \d^i_k \d(y-z)$ for
$E''^{ai}(z)$) in the right-hand side (\ref{eq 2varo}). Note that
these objects have the right geometric properties. We find
\bea \lefteqn{\frac{\d ^2 \O}{\d E^{bj}(x) \d E^{c'k'}(y)}=}
\nonumber\\*[6pt]
& & -\frac{1}{2} \ee^{bef}\left\{ (\pa_j^{(x)} \d^{ec'}\d^m_{k'}+
\G'^m_{jk'}(x) \d^{ec'}
+\ee^{eac'}\o^{a}_j(x)\d^m_{k'})\d(x-y)\right\}E^f_m(x)
\nonumber\\*[6pt]
& & +\frac{1}{2} \ee^{def}\left\{(\pa_{\ell}^{(x)}\d^{ec'}
 \d^m_{k'}+\G'^m_{\ell k'}(x) \d^{ec'}
+\ee^{eac'}\o^{a}_{\ell}(x)\d^m_{k'})\d(x-y)
\right\} E^{d\ell}(x)E^b_m(x)E^f_j(x)
\nonumber\\*[6pt] & & \label{eq ech}+2
\mbox{ terms with } (b \leftrightarrow c') (j \leftrightarrow k')
(x \leftrightarrow y) \eea
The result is a distribution of order 1, and when we parallel-transport
it, we can expand
\bea M^{k'c c'}_k(x,y)& =& M^{k'c c'}_k(x,x) +(y^{\ell}-x^{\ell})
\left(\pa_{\ell}^{(y)}M^{k'c c'}_k\right)(x,x) + \cdots
\nonumber\\*[8pt]
& = &
\d^{k'}_k\d^{cc'}-(y^{\ell}-x^{\ell}) (\G^{k'}_{\ell k}(x) \d^{cc'}
+\o^{cc'}_{\ell}(x) \d^{k'}_k) +\cdots \eea
where the missing terms annihilate $\d(x-y)$ and its first derivative,
and consequently do not contribute. One should also expand the electric
field in (\ref{eq ech}) as
\bea E^{ai}(y)& = & E^{ai}(x)+(y^{\ell}-x^{\ell})(\pa_{\ell} E^{ai})(x)
+\cdots
\nonumber\\*[8pt]
& = & E^{ai}(x)+(y^{\ell}-x^{\ell})(-\G'^i_{\ell m} E^{am}
+\o^{ad}_{\ell} E^{di}) + \cdots \eea
(and the corresponding equation for $E^a_i$), so that the evaluation
point is always $x$. Then all that remains is a lengthy but
straightforward computation. All the terms involving $\o$ cancel
either because of the antisymmetry of the structure constants of
$SU(2)$ or because of the Jacobi identity. The terms involving $\G$
correspond to those involving $\G'$ with opposite signs, so that the
final result is
\bea \lefteqn{\left[\frac{\d ^2 \O}{\d E^{bj}(x) \d E^{ck}(y)}
\right]^{(cov)}= }
\nonumber\\*[6pt] & & -\frac{1}{2}\ee^{def}e^f_m(x) \left\{
\d^{bd} \d^{ce} \d^{\ell}_j \d^m_k - \d^{cd} \d^{be} \d^{\ell}_k
\d^m_j +(\d^{bd} \d^m_k e^c_j(x) -\d^{cd} \d^m_j e^b_k(x)) e^{e\ell}(x)
\right\}\pa^{(x)}_{\ell}\frac{\d (x-y)}{\sqrt G} \eea

This is manifestly a tensorial object, and it is antisymmetric under
the simultaneous exchange $ (b \leftrightarrow c) (j \leftrightarrow
k)$. Hence the contraction with $\ee ^{ijk}\ee ^{abc}$ vanishes
identically, and the regulated version of
\beq \ee ^{ijk}\ee ^{abc} \frac{\d \o^c_k(x)}{\d E^{bj}(x)} \eeq
vanishes as announced above.

\end{document}